\documentclass[prx,a4paper,twocolumn,10pt,floatfix,superscriptaddress,amsmath,amsfonts,amssymb,preprintnumbers,citeautoscript,aps,longbibliography]{revtex4-2}

\usepackage{graphicx}
\usepackage{dcolumn}
\usepackage{bm}
\usepackage[colorlinks,plainpages=false,linkcolor=blue,urlcolor=blue,citecolor=blue,pdfpagemode=UseNone,pdfstartview=FitBH]{hyperref}
\usepackage{ulem}

\usepackage[english]{babel}
\usepackage[utf8]{inputenc}
\usepackage{mathtools,mathrsfs}
\usepackage{relsize}
\usepackage{mdframed}
\usepackage[margin=0.6in]{geometry}
\usepackage{enumitem}
\usepackage{float}
\usepackage{url}
\usepackage[table,usenames,dvipsnames]{xcolor}
\usepackage{gensymb}
\usepackage{grffile}
\usepackage[bottom]{footmisc}
\usepackage[capitalize]{cleveref}
\usepackage{textcomp}
\usepackage{braket} 

\makeatletter
\def\@fnsymbol#1{\ensuremath{\ifcase#1\or \dagger\or \ddagger\or  
*\or \mathsection\or \mathparagraph\or \|\or **\or \dagger\dagger
\or \ddagger\ddagger \else\@ctrerr\fi}}
\makeatother

\begin{document}

\title{Spin-liquid-like ground states in the double hydroxyperovskites\\CuSn(OD)$_6$ and MnSn(OD)$_6$ evidenced by $\mu$SR spectroscopy}
\author{Moumita Naskar}\email[E-mail: ]{moumita.naskar@tu-dresden.de}
\address{Institut f{\"u}r Festk{\"o}rper und Materialphysik, Technische Universit{\"a}t Dresden, 01062 Dresden, Germany} 
\author{Anton A. Kulbakov}
\address{Institut f{\"u}r Festk{\"o}rper und Materialphysik, Technische Universit{\"a}t Dresden, 01062 Dresden, Germany}
\author{Kaushick K. Parui}
\address{Institut f{\"u}r Festk{\"o}rper und Materialphysik, Technische Universit{\"a}t Dresden, 01062 Dresden, Germany}
\author{Jonas A. Krieger}
\address{PSI Center for Neutron and Muon Sciences, 5232 Villigen PSI, Switzerland}
\author{Thomas J. Hicken}
\address{PSI Center for Neutron and Muon Sciences, 5232 Villigen PSI, Switzerland}
\author{Hubertus Luetkens}
\address{PSI Center for Neutron and Muon Sciences, 5232 Villigen PSI, Switzerland}
\author{Ellen H{\"a}u$\ss$ler}
\address{Fakult{\"a}t f{\"u}r Chemie und Lebensmittelchemie, Technische Universit{\"a}t Dresden, 01062 Dresden, Germany}
\author{Thomas Doert}
\address{Fakult{\"a}t f{\"u}r Chemie und Lebensmittelchemie, Technische Universit{\"a}t Dresden, 01062 Dresden, Germany}
\author{Darren C. Peets}
\address{Institut f{\"u}r Festk{\"o}rper und Materialphysik, Technische Universit{\"a}t Dresden, 01062 Dresden, Germany}
\author{Hans-Henning Klauss}
\address{Institut f{\"u}r Festk{\"o}rper und Materialphysik, Technische Universit{\"a}t Dresden, 01062 Dresden, Germany}
\author{Dmytro S. Inosov}\email[E-mail: ]{dmytro.inosov@tu-dresden.de}
\address{Institut f{\"u}r Festk{\"o}rper und Materialphysik, Technische Universit{\"a}t Dresden, 01062 Dresden, Germany}
\address{W{\"u}rzburg-Dresden Cluster of Excellence on Complexity and Topology in Quantum Matter — ct.qmat, Technische Universität Dresden, 01062 Dresden, Germany}
\author{Rajib Sarkar}\email[E-mail: ]{rajib.sarkar@tu-dresden.de}
\address{Institut f{\"u}r Festk{\"o}rper und Materialphysik, Technische Universit{\"a}t Dresden, 01062 Dresden, Germany}

\begin{abstract}\noindent
Double hydroxide perovskites with magnetic transition-metal ions were recently identified as a unique class of materials that combine magnetic frustration with correlated proton disorder\,---\,a prerequisite for quantum-disordered fluctuating magnetic ground states resembling spin liquids. Here we present the results of muon spin relaxation ($\mu$SR) measurements carried out on fully deuterated samples of the double hydroxyperovskites CuSn(OH)$_6$ ($S=1/2$) and MnSn(OH)$_6$ ($S=5/2$) over the temperature range 0.053--50~K. The absence of any long-range magnetic order is confirmed down to 0.053~K. We observe no oscillations of the muon asymmetry down to the lowest temperature. The muon relaxation rates show a continuous increase with decreasing temperature, indicating persistent spin fluctuations in both compounds. Spin correlations are consistent with homogeneous spin dynamics. These observations reinforce the assertion that both compounds have a quantum-dynamic magnetic ground state that is consistent with a spin-liquid-like phase stabilized by proton disorder.
\end{abstract}

\maketitle

\paragraph*{Introduction} Frustrated magnetism is a forefront topic in condensed matter physics, providing a platform to explore emergent phenomena and exotic quantum phases \cite{Wolf2001, Lee2008, Balents2010, Savary2017, Zhou2017, Liu2018}. Magnetic frustration, arising from competing interactions, can suppress long-range order at low temperatures and can form a quantum-entangled ground state without long-range order, a so-called quantum spin liquid (QSL) \cite{Lee2008, Balents2010, Savary2017, Balents2010}. The concept of a quantum spin liquid was introduced by Anderson in 1973 as a resonating valence bond (RVB) ground state of interacting Heisenberg spins on a two-dimensional triangular-lattice antiferromagnet~\cite{Anderson1973}. Although no material has yet been universally accepted as a definitive realization of a QSL, a number of systems have been reported to show properties consistent with such a state~\cite{Shimizu2003, Helton2007, Yamashita2010, Zhou2011, Khuntia2016, Yuesheng2016, Banerjee2017, Baenitz2018, Sarkar2019, Clark2019, Bachhar2024}. However, in many cases, quenched disorder originating from site mixing, atomic substitutions, or lattice defects can produce thermodynamic and $\mu$SR signatures that closely mimic those expected for a QSL~\cite{Freedman2010, Calder2013, Zhu2017, Zorko2017, Murayama2020, Smaha2020, Hu2021, Sana2024}. Understanding how this local randomness influences and modifies intrinsic quantum fluctuations remains a key challenge in this field.

Hydroxyperovskites are a class of transition-metal compounds that form perovskite-related crystal structures derived from the ReO$_3$-type aristotype, also known as the $A$-site-vacant perovskite structure~\cite{Bock2002}. It consists of corner-sharing oxygen octahedra centered on the transition-metal ions that occupy the $B$ site, which are arranged in a distorted perovskite-like lattice in which protons play the role of the missing $A$-site cations~\cite{Basciano1998, Lafuente2015, Evans2020, Kampf2024, Welch2025}. Many of these compounds occur as rare natural minerals (for reviews, see Refs.~\cite{Mitchell2017, Welch2025}). In particular, the stannates CuSn(OH)$_6$ and MnSn(OH)$_6$ studied in this work correspond to the schoenfliesite-subgroup minerals mushistonite~\cite{MorgensternBadarau1976} and tetrawickmanite~\cite{White1973, Lafuente2015}, respectively.

Hydroxyperovskites exhibit correlated proton disorder~\cite{Bernal1933, Pauling1935, Malenkov2009}, which means that each hydrogen (or deuterium) ion statistically occupies one of two (or more) possible crystallographic sites, resulting in partial site occupancies. Such disorder can be tuned by hydrostatic pressure \cite{Kulbakov2025a}, potentially resulting in a series of proton-ordering transitions, analogous to those observed in conventional water ice.

\begin{figure}[b]
\includegraphics[width=\linewidth]{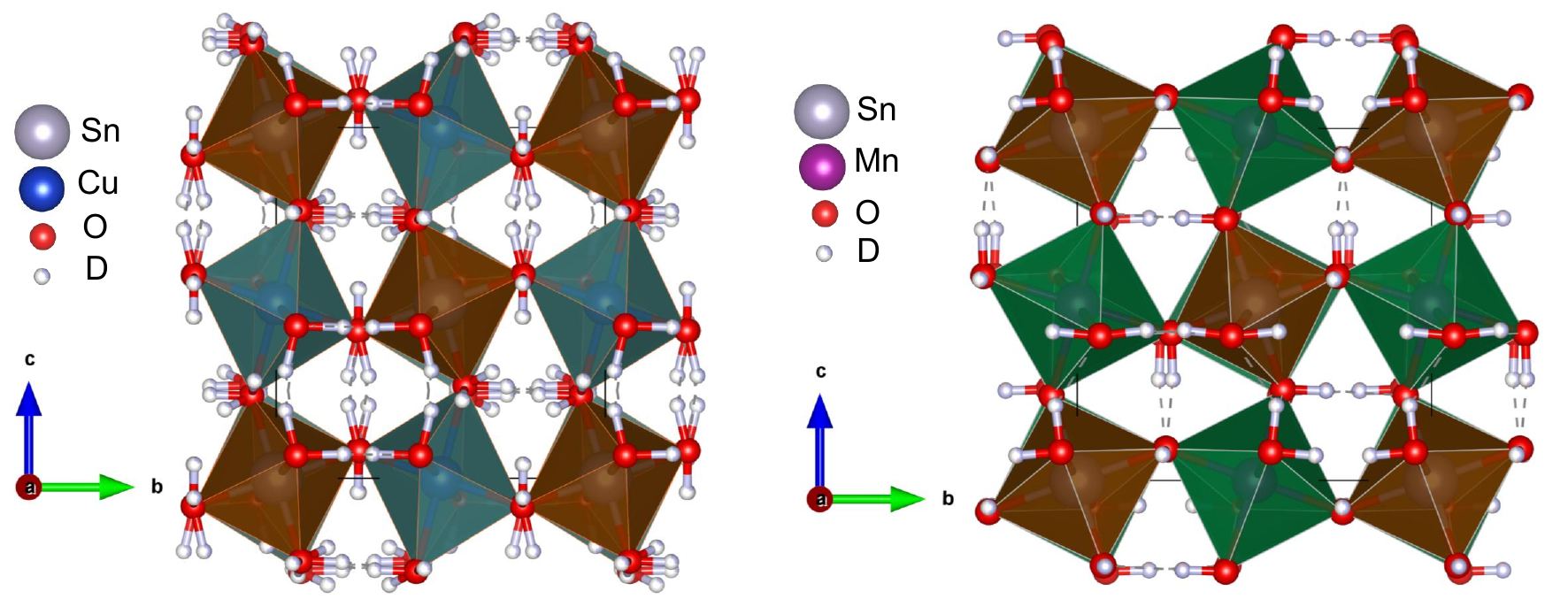}
\caption{Crystal structures of (a)~CuSn(OD)$_6$ with orthorhombic $Pnnn$ symmetry~\cite{Kulbakov2025a} and (b)~MnSn(OD)$_6$ with tetragonal $P4_2/n$ symmetry~\cite{Parui2025}. The Sn$^{4+}$-centered octahedra are shown in brown; the Cu$^{2+}$- and Mn$^{2+}$-centered octahedra in blue and green, respectively.}
\label{fig1}
\end{figure}

\begin{figure*}
	\includegraphics[width=\linewidth]{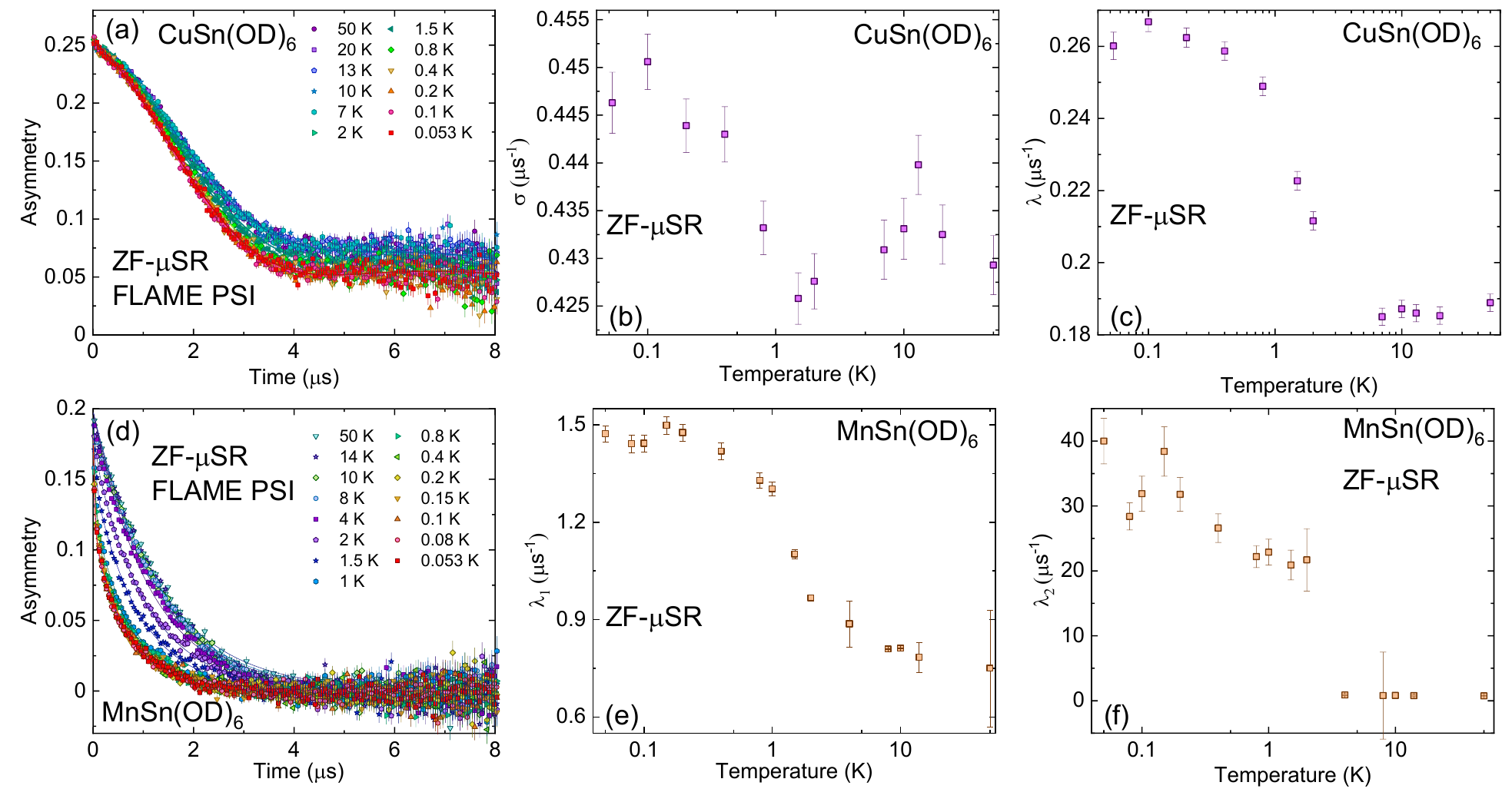}
	\caption{(a), (d) ZF-$\mu$SR time spectra measured at various temperatures for CuSn(OD)$_6$ and MnSn(OD)$_6$, respectively. Lines indicate the theoretical description as detailed in the text. (b), (c) Temperature dependencies of the width of the static field distribution and muon relaxation rate in zero-field $\mu$SR measurements of CuSn(OD)$_6$, respectively. (e), (f) Temperature evolution of the muon relaxation rates in ZF-$\mu$SR spectra of MnSn(OD)$_6$.}
   \label{fig2}
\end{figure*}
 
Just like in double perovskites, an ordered arrangement of two distinct cations on the B site results in the double hydroxyperovskite structure, in which either one or both cations can be magnetic. CuSn(OH)$_6$ and MnSn(OH)$_6$ are double hydroxyperovskites with one magnetic transition-metal ion (Cu$^{2+}$, $S=1/2$, or Mn$^{2+}$, $S=5/2$) and nominally nonmagnetic Sn$^{4+}$ (see Fig.~\ref{fig1}). The magnetic ions form a distorted fcc sublattice, which is geometrically frustrated and was theoretically considered, among others, as a potential host for spin liquids~\cite{Revelli2019, Kiese2022, Schick2022}. In CuSn(OH)$_6$, strong Jahn–Teller distortions of the Cu$^{2+}$ environment~\cite{MorgensternBadarau1976} partially relieve this frustration, yet no long-range magnetic order was observed down to 45~mK in earlier measurements performed on fully deuterated powder samples of CuSn(OD)$_6$~\cite{Kulbakov2025a, Kulbakov2025}. Similarly, MnSn(OD)$_6$ does not show long-range magnetic order down to 20~mK but instead exhibits neutron-scattering signatures of short-range, possibly dynamic, magnetic correlations at low temperatures~\cite{Parui2025}. In both compounds, broad anomalies in the specific heat and magnetization also indicate short-range dynamic spin correlations. Overall, CuSn(OD)$_6$ and MnSn(OD)$_6$ demonstrate spin-liquid mimicry governed by structural disorder in the proton sublattice~\cite{Kulbakov2025}. The deuteration serves here to minimize the incoherent scattering background in neutron-diffraction experiments on the same sample~\cite{Kulbakov2025a, Kulbakov2025, Parui2025}, otherwise we have not detected any significant differences in the magnetic properties of the protonated and deuterated samples.

In this study, we investigate CuSn(OD)$_6$ and MnSn(OD)$_6$ to understand the nature of their static and/or dynamic magnetic ground states. 
We carried out detailed $\mu$SR measurements in both zero field (ZF) and longitudinal field (LF) applied along the initial muon polarization, over a temperature range of 0.053--50 K. Being a local probe, $\mu$SR is highly sensitive to spin dynamics and can detect the presence or absence of tiny static internal fields.

The present $\mu$SR study evidences the absence of long-range magnetic order, together with the presence of persistent fluctuations down to 0.053~K in both compounds. This is consistent with a dynamically disordered ground state that mimics spin-liquid behavior.

\paragraph*{Experiment} Details of the synthesis and physical characterization of the CuSn(OD)$_6$ and MnSn(OD)$_6$ samples are provided in Refs.~\cite{Kulbakov2025a, Kulbakov2025, Parui2025}. The $\mu$SR experiments were performed at the Paul Scherrer Institute, Switzerland, using the FLAME instrument. For measurements, a pellet was prepared by pressing 0.48~g of CuSn(OD)$_6$ powder mixed with six drops of a diluted GE varnish--alcohol mixture; a second pellet was prepared from 0.49~g of MnSn(OD)$_6$ powder using Apiezon~N grease as a binding medium. Both pellets had a diameter of 13~mm and a thickness of 1~mm. The samples were mounted by sandwiching each pellet between two 25~$\mu$m copper foils, which were loaded into a Kelvinox dilution refrigerator insert for the measurement. The $\mu$SR data were analyzed with the \textsc{Musrfit} program~\cite{Suter2012}. Figures~\ref{fig1}(a) and \ref{fig1}(b) depict the crystal structures of CuSn(OD)$_6$ and MnSn(OD)$_6$, drawn with \textsc{Vesta}~\cite{Momma2011}. CuSn(OD)$_6$ crystallizes in an orthorhombic $Pnnn$ structure with alternating [Cu$^{2+}$(OD)$_6$] and [Sn$^{4+}$(OD)$_6$] octahedra, while MnSn(OD)$_6$ crystallizes in a tetragonal  $P4_2/n$ structure with alternating [Mn$^{2+}$(OD)$_6$] and [Sn$^{4+}$(OD)$_6$] octahedra.

\paragraph*{Results and discussion} Figs.~\ref{fig2}(a,\,d) show the time dependence of muon polarization at zero field for both compounds. They reveal no asymmetry oscillations down to 0.053~K, demonstrating the absence of a well-defined static internal field in either system.

\begin{figure*}
	\includegraphics[width=\linewidth]{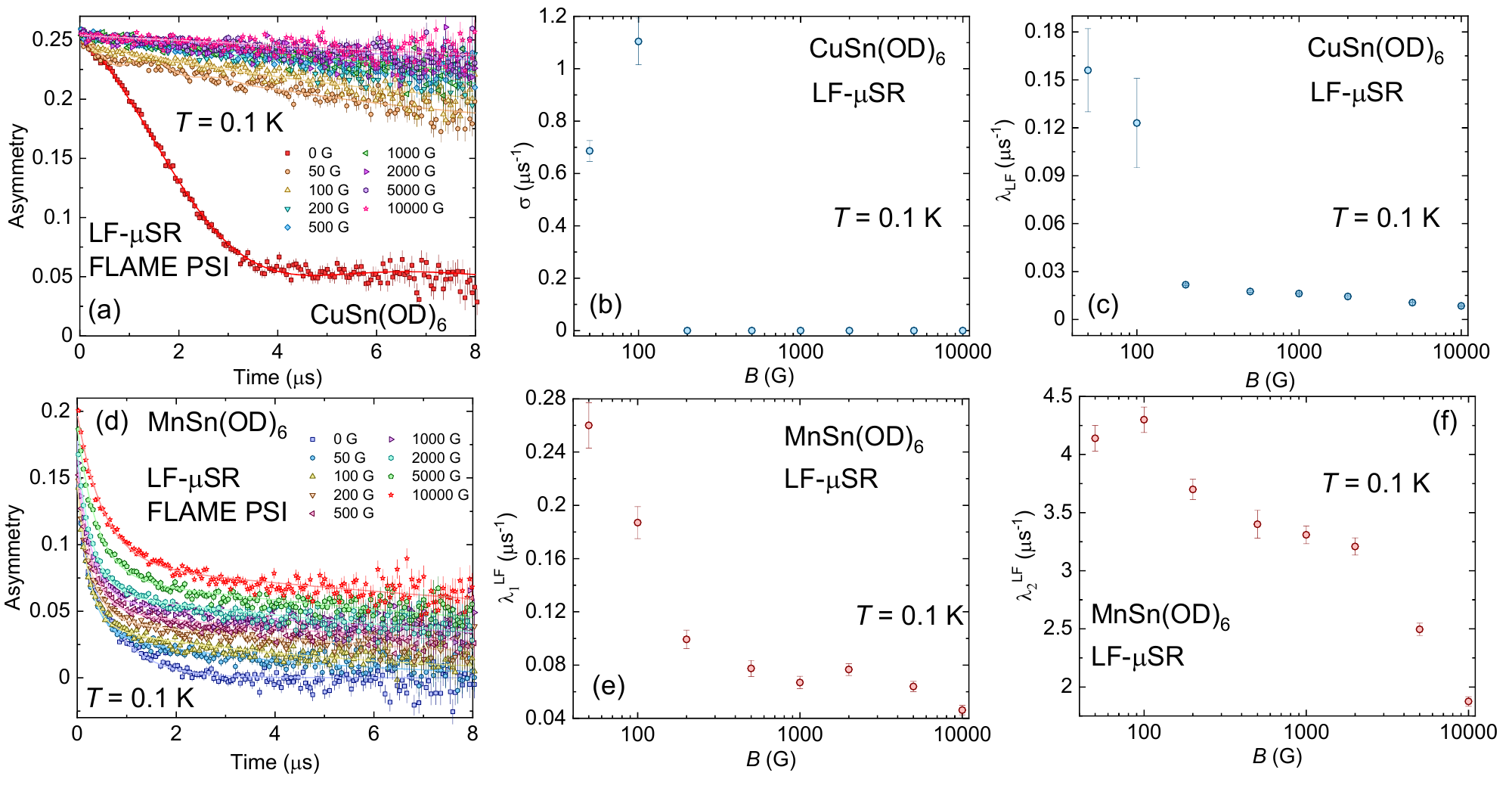}
	\caption{(a), (d) LF-$\mu$SR asymmetry spectra of CuSn(OD)$_6$ and MnSn(OD)$_6$ systems measured at 0.1 K. Lines correspond to the theoretical model discussed in the text. (b), (c) Field dependencies of the width of the static field distribution, muon relaxation rate in LF-$\mu$SR measurements of CuSn(OD)$_6$. (e), (f) Field evolution of the muon relaxation rates in LF-$\mu$SR spectra of MnSn(OD)$_6$.}
   \label{fig3}
\end{figure*}

A two-component model effectively describes the zero-field $\mu$SR data [Fig.~\ref{fig2}\,(a)] of CuSn(OD)$_6$ throughout the temperature range measured: a small static fraction and an exponential relaxation term. 
\begin{equation}
    A(t) = A_0[f G^{\rm KT} (t, \sigma)+(1-f)\exp(-\lambda t)], 
\end{equation}
where $A_0$ represents the initial asymmetry, $G^{\rm KT}$ is the static Gaussian Kubo-Toyabe (KT) relaxation function, $f$ is temperature-dependent fraction described by the KT function, and $\sigma$ and $\lambda$ are the width of the static field distribution and the muon relaxation rate, respectively. The KT term originates from an isotropic Gaussian distribution of randomly oriented static or quasi-static local fields while the exponential term with relaxation rate $\lambda$ often originates from spin fluctuations. The temperature dependence of $\sigma$ and $\lambda$ are presented in Figs.~\ref{fig2}(b) and \ref{fig2}(c), respectively. 
The variation of $\sigma$ with temperature indicates the onset of quasi-static spin correlations in the low-temperature region. With decreasing temperature, $\lambda$ increases monotonically and becomes almost constant below 0.2 K, while above 7 K it remains nearly temperature-independent, consistent with a paramagnetic region dominated by short-time spin fluctuations. 
In the intermediate temperature range 0.4 K $<T<$ 7 K, a significant enhancement in the $\mu$SR relaxation rate is observed on cooling, supporting further development of spin correlations at low temperatures.
Such features are consistent with other spin-liquid candidates investigated with muons \cite{Uemura1994, Clark2013, Li2016, Sarkar2019, Ding2019}.

The ZF-$\mu$SR asymmetry of MnSn(OD)$_6$ exhibits an exponential decay in the measured temperature range [Fig.~\ref{fig2}\,(d)]. We analyze these data by fitting the muon spectra with two exponential components:
\begin{equation}\label{MnEqn}
    A(t) = A_1[f_1\exp(-\lambda_1t)+(1-f_1)\exp(-\lambda_2t)], 
\end{equation}
where $A_1$ denotes the initial asymmetry, $\lambda_1$ and $\lambda_2$ represent the muon spin relaxation rates for muons likely implanted at two different positions in the lattice, and $f_1$ stands for the fraction which is temperature dependent following $\lambda_1$. That the muon time spectra are best described by an exponential decay function indicates that muon spin relaxation is primarily caused by spin fluctuations.
Figures~\ref{fig2}(e,\,f) show the temperature dependence of the relaxation rates. Below 0.2~K, $\lambda_1$ tends to saturate [Fig.~\ref{fig2}\,(e)], indicating invariant spin dynamics and a correlated disordered state. $\lambda_2$ [Fig.~\ref{fig2}\,(f)] also increases on cooling, but the scatter in the points limits what can be said about the form of this increase. The behavior found here is in line with other reported data, such as magnetic, thermodynamic, and neutron data \cite{Parui2025}. 

Above 7~K, $\lambda$ (= 1/$T_\text{1}^\mu$) is almost temperature independent in both materials. This indicates that it is likely dominated by exchange fluctuations of the Cu and Mn spins in CuSn(OD)$_6$ and MnSn(OD)$_6$. The exchange interaction ($J$) can be estimated using $k_\text{B}\theta_\text{CW}$ = $\frac{1}{3}zS(S+1)J$, where $z$ = 12 corresponds to the coordination number of the nearly cubic environment in both compounds. Using $S = 1/2$,  $\theta_\text{CW} = -7.1$~K for CuSn(OD)$_6$ \cite{Kulbakov2025} and $S = 5/2$, $\theta_\text{CW} = -7.34$~K for MnSn(OD)$_6$~\cite{Parui2025}, we obtain the spin fluctuation rates $\nu$ using~\cite{Uemura1994} $\nu = \sqrt{z}JS/h$ $\approx 85.3$~GHz $\approx 0.35$~meV for CuSn(OD)$_6$ and $\nu\approx 37.8$~GHz $\approx 0.16$~meV for MnSn(OD)$_6$. Using the zero-field relation $\lambda = 2\Delta^2/\nu$, the high-temperature internal field distribution widths $\Delta$ are estimated to be $89.3$~$\mu$s$^{-1}$ for CuSn(OD)$_6$ and 123.7~$\mu$s$^{-1}$ for MnSn(OD)$_6$, which are much smaller than the fluctuation rates, $\nu$. These results are consistent with the muon spin relaxation being in the fast fluctuation limit \cite{Uemura1994, Bono2004, Li2016, Sarkar2019}.

Having discussed the ZF $\mu$SR spectra, we now proceed to the LF-$\mu$SR results for both compounds. Figures~\ref{fig3}(a,\,d) present the field-dependent $\mu$SR time spectra measured at 0.1~K in longitudinal field for both compounds. For CuSn(OD)$_6$, the static or quasistatic component is easily decoupled by a weak longitudinal field [Fig.~\ref{fig3}\,(a)]. The LF-$\mu$SR time spectra of CuSn(OD)$_6$ can be well described using two components: a static Gaussian Kubo–Toyabe relaxation in a nonzero longitudinal field \cite{Hayano1979} and an exponential relaxation term:
\begin{equation}
    A(t) = A_2[f_2 G^{\rm KT} (\nu, \sigma, t, H_{\rm LF})+(1-f_2)\exp(-\lambda_\text{LF}t)] .
\end{equation}
For longitudinal field $H_\text{LF} = 100$~G [Fig.~\ref{fig3}\,(a)] the static contributions to the spectra are nearly completely decoupled and the relaxation is mainly dynamic. Upon fitting the data above 100~G, we found that $\sigma$ comes out to be zero, and the $\mu$SR time relaxation spectra follow a single exponential decay:
\begin{equation}
    A(t) = A_2 \exp(-\lambda_\text{LF}t).
\end{equation}
To ensure a consistent analysis and to avoid unnecessary complications in the behavior of the relaxation rate ($\lambda_{\mathrm{LF}}$), we fixed $\sigma = 0$ and the fraction $f_2 = 0$ for longitudinal fields above 100~G during the fitting procedure. This choice is justified by the fact that the application of a 100~G longitudinal field partially decouples the system from the Gaussian relaxation component, which most likely originates from static nuclear contributions. Consequently, this approach allows us to distinctly quantify the longitudinal muon relaxation rates, as qualitatively evident from Fig.~\ref{fig3}(a).

 The exponential form of the LF-$\mu$SR relaxation suggests that the spin dynamics are possibly similar throughout the sample. In Fig.~\ref{fig3}\,(c), we show the decay rate ($\lambda_\text{LF}$) at $T = 0.1$~K extracted from these fits, as a function of the magnetic field. The pronounced decrease in $\lambda_\text{LF}$ above 100 Oe is due to the decoupling of the nuclear contribution to the relaxation rate. Beyond this field, $\lambda_\text{LF}$, now reflecting only the electronic contribution, remains nearly unchanged up to 1 T. This behavior indicates that the spin correlations are likely to be homogeneous spin dynamics rather than inhomogeneous distributions, as expected for a quantum spin liquid state \cite{Tripathi2022, Bhattacharya2024}.

In MnSn(OD)$_6$, the muon spin depolarization [Fig.~\ref{fig3}\,(d)] is weakly affected by a longitudinal field. Even 1-T LF is not sufficient to suppress the muon relaxation at 0.1 K completely, partially decoupling only a fraction of the signal in the tail region. This means the magnetic ground state is dynamic at the base temperature. The spectra were fitted with Eq.~\ref{MnEqn}, using $\lambda^\text{LF}_1$ and $\lambda^\text{LF}_2$ to represent the muon spin relaxation rates with longitudinal fields applied. As shown in Figs.~\ref{fig3}(e,\,f), the decreasing value of these relaxation rates on warming indicates a gradual decoupling of the muon spin polarization from the sample.

As we can see comparing the fitting results for both compounds, different fitting models are required, likely reflecting differences in the crystallographic and magnetic environments of the muon stopping sites and the longitudinal-field strengths needed for decoupling. Nevertheless, for both compounds the low-temperature relaxation data in zero field and under longitudinal field are consistent with a fluctuating, homogeneous ground state, supporting their identification as quantum spin-liquid candidates \cite{Clark2013, Li2016, Sarkar2019}. Understanding the origin of the different relaxation mechanisms will require a detailed determination of the muon stopping sites, including precise Knight-shift calculations.

One striking observation is that, for the CuSn(OD)$_6$ system, a longitudinal field of only 100\,G is sufficient to decouple the relaxation from nuclear contributions, whereas in MnSn(OD)$_6$ the nuclear contribution appears to be weak or negligible, requiring substantially larger longitudinal fields for decoupling. Based on this observation alone, it is not straightforward to argue that one compound is more dynamic than the other. Instead, for both systems the low-temperature behavior is consistent with quantum spin-liquid physics \cite{Li2016, Sarkar2019, Clark2019, Pula2024}.
\paragraph*{Conclusions} In conclusion, detailed $\mu$SR studies on the CuSn(OD)$_6$ and MnSn(OD)$_6$ powder samples have been carried out. No evidence of long-range magnetic order is observed in either sample down to 0.053~K. Below 0.2~K, the $\mu$SR relaxation rate remains nearly constant in both compounds, suggesting a disordered ground state. From the $\mu$SR point of view, CuSn(OD)$_6$ and MnSn(OD)$_6$ appear to be promising candidates for a spin-liquid-like ground state. The present results do not give any direct evidence about the proton disorder in these compounds; to explore the proton disorder further, local-probe studies such as nuclear magnetic resonance (NMR) are needed.


\begin{acknowledgments} We thank M.~D. Welch from the Natural History Museum, London, for fruitful discussions regarding the crystal structures of the hydroxyperovskites. This project was funded by the Deutsche Forschungsgemeinschaft (DFG, German Research Foundation) through individual research Grants \mbox{IN 209/12-1} and \mbox{DO 590/11-1} (Project No.~536621965); projects B03, C02, C03, and C06 of the Collaborative Research Center SFB 1143 (Project No.~247310070); and the Würzburg-Dresden Cluster of Excellence on Complexity and Topology in Quantum Materials—\textit{ct.qmat} (EXC~2147, Project No.~390858490). M.\,N. acknowledges financial support from the Deutsche Forschungsgemeinschaft via the Walter Benjamin Program (Grant NA 2012/1-1, Project No.~540160192). This work is based on experiments performed at the Swiss Muon Source S$\mu$S, Paul Scherrer Institute, Villigen, Switzerland.
\end{acknowledgments}

\bibliographystyle{apsrev4-2}
\bibliography{uSR-reference}

\end{document}